\documentclass[twocolumn]{revtex4}

\usepackage{graphicx}
\usepackage{amsmath}
\usepackage{xcolor, color, framed}
\usepackage[colorlinks, linkcolor=red,anchorcolor=green,citecolor=blue]{hyperref}
\usepackage[normalem]{ulem}

\begin{document}
\title{Heavy flavor dissociation in the framework of multi-body Dirac equations}
\date{\today}
\author{Shuzhe Shi$^{a,b}$}
\email{shuzhe.shi@mcgill.ca}
\author{Jiaxing Zhao$^{a,b}$}
\author{Pengfei Zhuang$^a$}
\email{zhuangpf@mail.tsinghua.edu.cn}
\address{$^a$Physics Department, Tsinghua University, Beijing 100084, China\\
      	 $^b$Department of Physics, McGill University, Montreal, QC H3A 2T8, Canada}

\begin{abstract}
We study heavy flavor properties at finite temperature in the framework of a relativistic potential model. With an improved method to solve the three-body Dirac equation, we determine a universal set of model parameters for both mesons and baryons by fitting heavy flavor masses in vacuum. Taking heavy quark potential from lattice QCD simulations in hot medium, we systematically calculate heavy flavor binding energies and averaged sizes as functions of temperature. The meson and baryons are separately sequentially dissociated in the quark-gluon plasma, and the mesons can survive at higher temperature due to the stronger potential between quark-antiquark pairs than that between quark-quark pairs.   
\end{abstract}
\maketitle

\section{Introduction}
\label{s1}
Heavy quark masses are much larger than the typical temperature of the fireball formed in high energy nuclear collisions at Relativistic Heavy Ion Collider (RHIC) and Large Hadron Collider (LHC) and the typical energy scale of the Quantum Chromodynamics (QCD), $m_Q\sim (1-5)$ GeV $\gg T\sim (0.3-0.5)$ GeV and $m_Q \gg \Lambda_{QCD}$, their thermal production in the fireball can then be safely neglected and they are entirely originated from the initial hard processes which can be solidly calculated in the perturbation theory. On the way out of the fireball, the heavy quarks pass through the quark-gluon plasma (QGP) created in the collisions and strongly interact with it. Since heavy flavor hadrons inherit the information of the QGP carried by the heavy quarks, they have long been considered as a sensitive probe of the QGP formation~\cite{Matsui:1986dk}.    

For closed heavy flavors, one can neglect the parton creation-annihilation fluctuations and describe their static properties with non-relativistic potential models~\cite{isgur}. The study is directly extended from vacuum to hot medium, by taking heavy quark potential from lattice QCD at finite temperature~\cite{Petreczky:2010yn,Kaczmarek:1900zz}. Solving the two-body Schr\"odinger equation for a $Q\bar Q$ system, people obtained the sequential dissociation temperatures for charmonium and bottominum states~\cite{Satz:2005hx,Burnier:2015tda}. The potential model is also extended to studying open heavy flavor states like $D$ mesons and $\Xi_{cc}$ and $\Omega_{ccc}$ baryons, by using the two-body Dirac equation~\cite{Crater:2002fq,Crater:2008rt,Guo:2012hx,Shi:2013rga} or three-body Schr\"odinger equation~\cite{SilvestreBrac:1996bg,He:2014tga,Zhao:2016ccp,Zhao:2017gpq}. 

For heavy flavor baryons with a typical radius $0.5$~fm, the uncertainty principle leads to a characteristic momentum around $0.4$~GeV, corresponding to a relative velocity $v\sim c$ between any two constituents. Therefore, the relativistic effect should seriously be taken into account, especially for singly charmed baryons. On the other hand, the spin interaction, which results in the fine structure of hadrons and becomes more important in multi-quark states, is a relativistic effect and should self-consistently be included through the covariant Dirac equation. In this paper, we study the three-body Dirac equation for heavy flavor baryons in hot medium and determine the dissociation temperatures for all the heavy flavor hadrons in a self-consistent way.  

The covariant wave equation proposed by Sazdjian~\cite{Sazdjian:1988be,Sazdjian:1986aw} provides a way to find relativistic bound-state wave functions. The interaction potential and the wave function of the bound-state are related in a definite way to the kernel and the wave function of the Bethe-Salpeter equation~\cite{Salpeter:1951sz,Sazdjian:1986aw}. Crater and Alstine developed the two-body Dirac equation~\cite{Crater:2002fq,Crater:1983ew,Crater:1987hm,Crater:2008rt} for mesons which covariantly converts the Dirac equation with 16 degrees of freedom to a Klein--Gordon-like equation with highly disentangled degrees of freedom. In this framework, one can clearly see the relativistic corrections to the non-relativistic potential model through the Darwin term and many spin interaction terms. Later on, the three-body Dirac equation for baryons was developed by Whitney and Crater~\cite{Whitney:2011aa}, which combines Sazdjian's $N$-body relativistic potential model with the two-body Dirac equation.

To systematically describe heavy flavor mesons and baryons in a relativistic potential model at finite temperature, we first improve the method to solve the three-body Dirac equation. We will expand the baryon wave functions in terms of the two-body spherical harmonic oscillator states, which guarantees the completeness and orthogonality of the expansion, increases the accuracy of the calculation, and can be used to study both the ground and excited baryon states. We will also take a universal set of model parameters for both mesons and baryons. This makes the calculation more self-consistent and predictable. With the improved method, we then solve the two- and three-body Dirac equations in hot medium, and determine the heavy flavor dissociation temperatures by calculating their binding energies and averaged radii.    

\section{Three-body Dirac equation}
\label{s2}
The two-body Dirac equation which is used to describe heavy flavor mesons is discussed in detail in Refs.~\cite{Crater:2002fq,Crater:2008rt} in vacuum and Refs.~\cite{Guo:2012hx, Shi:2013rga} at finite temperature. We will focus in this section on the three-body Dirac equation for heavy flavor baryons. In the framework firstly developed by Whitney {\it et al}~\cite{Whitney:2011aa} based on Sazdjian's $N$-body relativistic potential model~\cite{Sazdjian:1988be}, the baryon wave function $\Psi({\bf r}_1,{\bf r}_2,{\bf r}_3)$ is controlled by the Schr\"odinger-like equation,
\begin{equation}
\left[\sum_{i=1}^{3} {{\bf p}_i^2\over 2\epsilon_i} + \sum_{i<j}^3{\epsilon_i + \epsilon_j\over 2\epsilon_i \epsilon_j}{\cal V}_{ij}\right]\Psi =E\Psi,
\label{dirac}
\end{equation}
where ${\bf r}_i$ and ${\bf p}_i$ are the quark coordinate and momentum, $E=\sum_{ij}(\epsilon_j^2 - m_j^2)/(6\epsilon_i)$ is the energy eigenvalue related to the effective quark mass $\epsilon_i$ and vacuum quark mass $m_i$, and the baryon mass $M$ is determined by the coupled equations,
\begin{equation}
M=3\epsilon_i + \sum_{j\neq i}{m_i^2-m_j^2\over \epsilon_i+\epsilon_j}.
\label{mass}
\end{equation}
The interaction between (anti-)quarks is described by the relativistic potential ${\cal V}_{ij}({\bf r}_{ij})$\cite{Crater:2008rt},
\begin{eqnarray}
{\cal V}_{ij} &=& 2m_{ij}S+S^2+2\epsilon_{ij}A-A^2+\Phi_D\nonumber\\
&&+{\boldsymbol \sigma}_i\cdot{\boldsymbol \sigma}_j\Phi_{SS}+{\bf L}_{ij}\cdot({\boldsymbol \sigma}_i+{\boldsymbol \sigma}_j)\Phi_{SO}\nonumber\\
&&+({\boldsymbol \sigma}_i\cdot\hat{\bf r}_{ij})({\boldsymbol \sigma}_j\cdot\hat{\bf r}_{ij}){\bf L}_{ij}\cdot({\boldsymbol \sigma}_i+{\boldsymbol \sigma}_j)\Phi_{SOT}\nonumber\\
&&+{\bf L}_{ij}\cdot({\boldsymbol \sigma}_i-{\boldsymbol \sigma}_j)\Phi_{SOD}+i{\bf L}_{ij}\cdot({\boldsymbol \sigma}_i\times{\boldsymbol \sigma}_j)\Phi_{SOX}\nonumber\\
&&+(3({\boldsymbol \sigma}_i\cdot\hat{\bf r}_{ij})({\boldsymbol \sigma}_j\cdot\hat{\bf r}_{ij})-{\boldsymbol \sigma}_i\cdot{\boldsymbol \sigma}_j)\Phi_T
\label{potential}
\end{eqnarray}
with the mass and energy parameters $m_{ij}=m_im_j/(\epsilon_i + \epsilon_j)$ and $\epsilon_{ij}=((\epsilon_i + \epsilon_j)^2-m_i^2-m_j^2)/(2(\epsilon_i + \epsilon_j))$, relative coordinate ${\bf r}_{ij}={\bf r}_i-{\bf r}_j$, unit vector $\hat{\bf r}_{ij}={\bf r}_{ij}/|{\bf r}_{ij}|$, and orbital and spin angular momenta ${\bf L}_{ij}$ and ${\boldsymbol \sigma}_i$. The non-relativistic central potential $V_{q\bar q}(r)$ between quark and antiquark can be separated into two parts, $V_{q\bar q}(r) = A_{q\bar q}(r)+S_{q\bar q}(r)$ with $A_{q\bar q}(r) = -\alpha_{q\bar q}/r$ and $S_{q\bar q}(r) = \sigma_{q\bar q} r$, where $A_{q\bar q}$ and $S_{q\bar q}$ control, respectively, the behavior of the potential at short and long distances. In vacuum one usually takes the Cornell potential~\cite{Eichten:1978tg}, including a Coulomb part which dominates the wave function around $r=0$ and a linear part which leads to the quark confinement. For $V_{qq}$ between two quarks one can still take the separation, $V_{qq}(r) = A_{qq}(r)+S_{qq}(r)$ with $A_{qq}(r) = -\alpha_{qq}/r$ and $S_{qq}(r) = \sigma_{qq} r$. The two couplings $\alpha_{qq}$ and $\sigma_{qq}$ in quark-quark potential are in general very different from $\alpha_{q\bar q}$ and $\sigma_{q\bar q}$ in quark-antiquark potential. However, considering only one gluon exchange leads to $\alpha_{qq}=\alpha_{q\bar q}/2$~\cite{Schwartz:2013pla}, and the recent lattice simulation~\cite{Alexandrou:2002sn} shows $\sigma_{qq} \approx\sigma_{q\bar q}/2$. All the relativistic corrections in the potential (\ref{potential}), including the Darwin term $\Phi_D$, spin-spin coupling $\Phi_{SS}$, spin-orbital couplings $\Phi_{SO}, \Phi_{SOT}, \Phi_{SOD}$ and $\Phi_{SOX}$, and tensor coupling $\Phi_T$, are explicitly shown in Ref.\cite{Crater:2008rt}. Note that, any $\Phi$ is a function of the distance $|{\bf r}_{ij}|$, the dependence of the potential ${\cal V}_{ij}$ on the azimuthal angles is reflected in the coefficients in front of~$\Phi$.

Solving the baryon mass $M$ and wave function $\Psi$ from the Schr\"odinger-like equation (\ref{dirac}) is not a simple eigenstate problem, as the eigenvalue $E$ or $M$ appears also on the left hand side of the equation. We solve $M$ or the binding energy $\mathcal {E} = M-\widetilde M\ (\widetilde M=m_1+m_2+m_3)$ by employing iteration method. For a given value $M^{(n)}$ at the $n$-th step, we first obtain the corresponding effective quark mass $\epsilon_i^{(n)}$ from the coupled equations (\ref{mass}), and then solve the Schr\"odinger-like equation (\ref{dirac}) as an eigenstate problem with eigenvalue $E^{(n)}$. With sufficiently large number of iteration steps, one could obtain the baryon mass up to arbitrary accuracy.

To solve the Schr\"odinger-like equation for multi-body bound state problem, a usually used way is to expand the wave function in terms of known functions. While in some special cases, for instance the ground state, one can use variational method and expand the state in terms of Gaussian wave packets~\cite{Hiyama:2018ivm}, a general and systematic study including both ground and excited states should be carried out in a complete and orthogonal Hilbert space. We employ a numerical framework similar to Ref.~\cite{Capstick:1986bm} to treat the three-body bound state problem. We solve the baryon mass and wave function in the Hilbert space constructed by spherical harmonic oscillator states which are by definition complete and orthogonal. Considering the mass difference among the three quarks in a general baryon state, we take different constituent masses in constructing the spherical harmonic oscillator states. This is different from Ref.~\cite{Capstick:1986bm} where all the three constituents have the same mass. In such a framework we can study both the ground and excited states.

Like a two-body problem, we factorize the three-body motion into a center-of-mass motion and a relative motion. To this end, we introduce the coordinates 
\begin{eqnarray}
{\bf R} &=& (\epsilon_1{\bf r}_1+\epsilon_2{\bf r}_2+\epsilon_3{\bf r}_3)/(\epsilon_1+\epsilon_2+\epsilon_3),\nonumber\\
\boldsymbol {\rho} &=& \sqrt{\epsilon_1\epsilon_2/(\overline m(\epsilon_1+\epsilon_2))}({\bf r}_1-{\bf r}_2),\nonumber\\
\boldsymbol{\lambda} &=& \sqrt{\epsilon_3/(\overline m(\epsilon_1+\epsilon_2)(\epsilon_1+\epsilon_2+\epsilon_3))}\nonumber\\
&&\times(\epsilon_1({\bf r}_3-{\bf r}_1)+\epsilon_2({\bf r}_3-{\bf r}_2))
\end{eqnarray}
instead of ${\bf r}_1$, ${\bf r}_2$ and ${\bf r}_3$, where $\overline m$ is an arbitrary parameter which automatically disappears in the end and does not affect the result~\cite{1998FBS....25..199K}. Accordingly, we use the center-of-mass momentum ${\bf P}$ and relative momenta ${\bf p}$ and ${\bf q}$ instead of ${\bf p}_1$, ${\bf p}_2$ and ${\bf p}_3$. 

We now focus on the relative motion of the three-body problem which controls the inner structure of the bound state. We expand the relative wave function in terms of two-body spherical harmonic oscillator states. For a single spherical harmonic oscillator, its Schr\"odinger equation with potential $\mu\omega^2r^2/2$ can be straightforwardly solved with solutions $\Psi_{nlm}({\bf r}) = \psi_{nl}(r)Y_l^m(\theta,\phi)$ and $E_{nl} = (2n+l+3/2)\hbar\omega$, where $n,l,m$ are the principal, orbital and magnetic quantum numbers, and $Y_l^m$ and $\psi_{nl}$ are the angular and radial parts of the wave function. Note that the radial wave function $\psi_{nl}$ does not depend on the mass $\mu$ and frequency $\omega$ separately, it depends on the combined quantity $\alpha=\sqrt{\mu\omega/\hbar}$.

The two-body spherical harmonic oscillator states are defined as a direct product of two single spherical harmonic oscillator states,
\begin{equation}
\left| n_\rho l_\rho m_\rho n_\lambda l_\lambda m_\lambda \right> =\left|n_\rho l_\rho m_\rho\right>\left|n_\lambda l_\lambda m_\lambda\right>
\end{equation}
with the same scaling parameter $\alpha$ for the two oscillators.

The above defined two-body spherical harmonic oscillator states are the exact solutions of a three-body bound state problem with interaction potential
\begin{equation}
\Phi(|{\bf r}_{ij}|) = {\epsilon_i^2 \epsilon_j^2 \omega^2\over (\epsilon_i+\epsilon_j)(\epsilon_1+\epsilon_2+\epsilon_3)} r_{ij}^2.
\end{equation}

With the two-body spherical harmonic oscillator states obtained, we define the Hilbert space as a direct product of the flavor, spin and coordinate spaces,
\begin{equation}
\left| \Psi_{FSC} \right> = \left| F \right> \times \left| S \right> \times \left| n_\rho l_\rho m_\rho n_\lambda l_\lambda m_\lambda \right>.
\label{hilbert}
\end{equation}
The possible flavor and spin states are shown in Appendix \ref{aa}. In such a Hilbert space we expand the baryon wave function
\begin{equation}
\left| \Psi \right> = \sum_{FSC} C_{FSC} \left| \Psi_{FSC} \right>.
\end{equation}
Correspondingly, the Hamiltonian operator $H$ of the system becomes a matrix in the Hilbert space,
\begin{equation}
H_{FSC,F'S'C'} =  \left< \Psi_{FSC} \right| H \left| \Psi_{F'S'C'} \right>.
\end{equation}

Taking into account the complete and orthogonal conditions for the states $\left| \Psi_{FSC}\right>$, the eigenstate problem of a three-body system, $H\left| \Psi \right> = E \left| \Psi \right>$, becomes a matrix equation for the coefficients $C_{FSC}$,
\begin{equation}
\sum_{F'S'C'} H_{FSC,F'S'C'} \, C_{F'S'C'} = E \, C_{FSC}
\end{equation}
with the Hamiltonian matrix elements 
\begin{eqnarray}
&& H_{FSC,F'S'C'}\\
&=& \int \mathrm{d}^3\boldsymbol{\rho}\,\mathrm{d}^3\boldsymbol{\lambda}\,\Psi_{n_\rho l_\rho m_\rho}^*(\boldsymbol{\rho})\Psi_{n_\lambda l_\lambda m_\lambda}^*(\boldsymbol{\lambda})\nonumber\\
&&\times\left< F \right|\left< S\right|K +W\left|S' \right> \left| F'\right>\Psi_{n'_\rho l'_\rho m'_\rho}(\boldsymbol{\rho}) \Psi_{n'_\lambda l'_\lambda m'_\lambda}(\boldsymbol{\lambda}).\nonumber
\end{eqnarray}
The calculation for the kinetic energy $K=\sum_i p_i^2/(2\epsilon_i)$, which is proportional to $\nabla_\rho^2+\nabla_\lambda^2$, is straightforward. The difficulty comes from the calculation for the potential part $W=\sum_{i<j} (\epsilon_i + \epsilon_j){\cal V}_{ij}({\bf r}_{ij})/(2\epsilon_i\epsilon_j)$. For the potential ${\cal V}_{12}({\bf r}_{12})$ between the first and second constituents, since ${\bf r}_{12}$ is proportional to $\boldsymbol{\rho}$ and independent of $\boldsymbol{\lambda}$, the six dimensional integration $\int  \mathrm{d}^3\boldsymbol{\rho}\, \mathrm{d}^3\boldsymbol{\lambda}$ is reduced to a simple radial integration $\int \rho^2 \mathrm{d}\rho$.
For the other two potentials ${\cal V}_{23}({\bf r}_{23})$ and ${\cal V}_{13}({\bf r}_{13})$, since ${\bf r}_{23}$ and ${\bf r}_{13}$ depend on both $\boldsymbol{\rho}$ and $\boldsymbol{\lambda}$, the integration over $\boldsymbol{\rho}$ and $\boldsymbol{\lambda}$ can not be so simplified. We then make a rotation in the coordinate space from $(\boldsymbol{\rho},\boldsymbol{\lambda})$ to $(\widetilde{\boldsymbol{\rho}},\widetilde{\boldsymbol{\lambda}})$ to guarantee ${\bf r}_{23}\sim \widetilde{\boldsymbol{\rho}}$ or ${\bf r}_{13}\sim \widetilde{\boldsymbol{\rho}}$. This rotation in coordinate space is equivalent to a particle index rotation from the ordering $\{1,2,3\}$ to $\{2,3,1\}$ or $\{3,1,2\}$. The rotation can be explicitly written as
\begin{eqnarray}
&&\Psi_{n_\rho l_\rho m_\rho}(\boldsymbol{\rho}) \Psi_{n_\lambda l_\lambda m_\lambda}(\boldsymbol{\lambda})\\
&=& \sum_{\widetilde n_\rho \widetilde l_\rho \widetilde m_\rho \widetilde n_\lambda \widetilde l_\lambda \widetilde m_\lambda} D_{\widetilde n_\rho \widetilde l_\rho \widetilde m_\rho \widetilde n_\lambda \widetilde l_\lambda \widetilde m_\lambda}^{n_\rho l_\rho m_\rho n_\lambda l_\lambda m_\lambda}\Psi_{\widetilde n_\rho \widetilde l_\rho \widetilde m_\rho}(\widetilde{\boldsymbol{\rho}})\Psi_{\widetilde n_\lambda \widetilde l_\lambda \widetilde m_\lambda}(\widetilde{\boldsymbol{\lambda}}),\nonumber
\label{rotation}
\end{eqnarray}
where $D$ is the rotation matrix. After the rotation, the six dimensional integration in calculating the expectation values of ${\cal V}_{23}$ and ${\cal V}_{13}$ becomes again a simple radial integration $\int {\widetilde\rho}^2 \mathrm{d}\widetilde\rho$. The Hamiltonian matrix elements are explicitly listed in Appendix \ref{aa}.  

\section{Heavy flavors in vacuum}
\label{s3}
Before we solve the two- and three-body Dirac equations at finite temperature, we first fix the model parameters by fitting the heavy flavor masses in vacuum. Note that, the completeness of the two-body oscillator states $\left| n_\rho l_\rho m_\rho n_\lambda l_\lambda m_\lambda \right>$ is independent of the value of the scaling parameter $\alpha$, if we take infinite number of states. When the number of the states is finite in practical computation, the accuracy of the calculation depends on the choice of the $\alpha$ value.

For a baryon ground state, the $\alpha$ value corresponding to the most fast convergence can be fixed by applying variational method. We take $\alpha$ as a parameter and calculate the energy eigenvalue $E$ and wave function $\Psi$ as functions of $\alpha$. By minimizing the eigenvalue
\begin{equation}
\label{variation}
{\partial E\over \partial \alpha}=0,\ \ \ \ \ \ {\partial^2 E\over \partial\alpha^2}>0,
\end{equation}
we obtain the $\alpha$ value corresponding to the most fast convergence.

\begin{figure}[!hbt]
	\includegraphics[width=200pt]{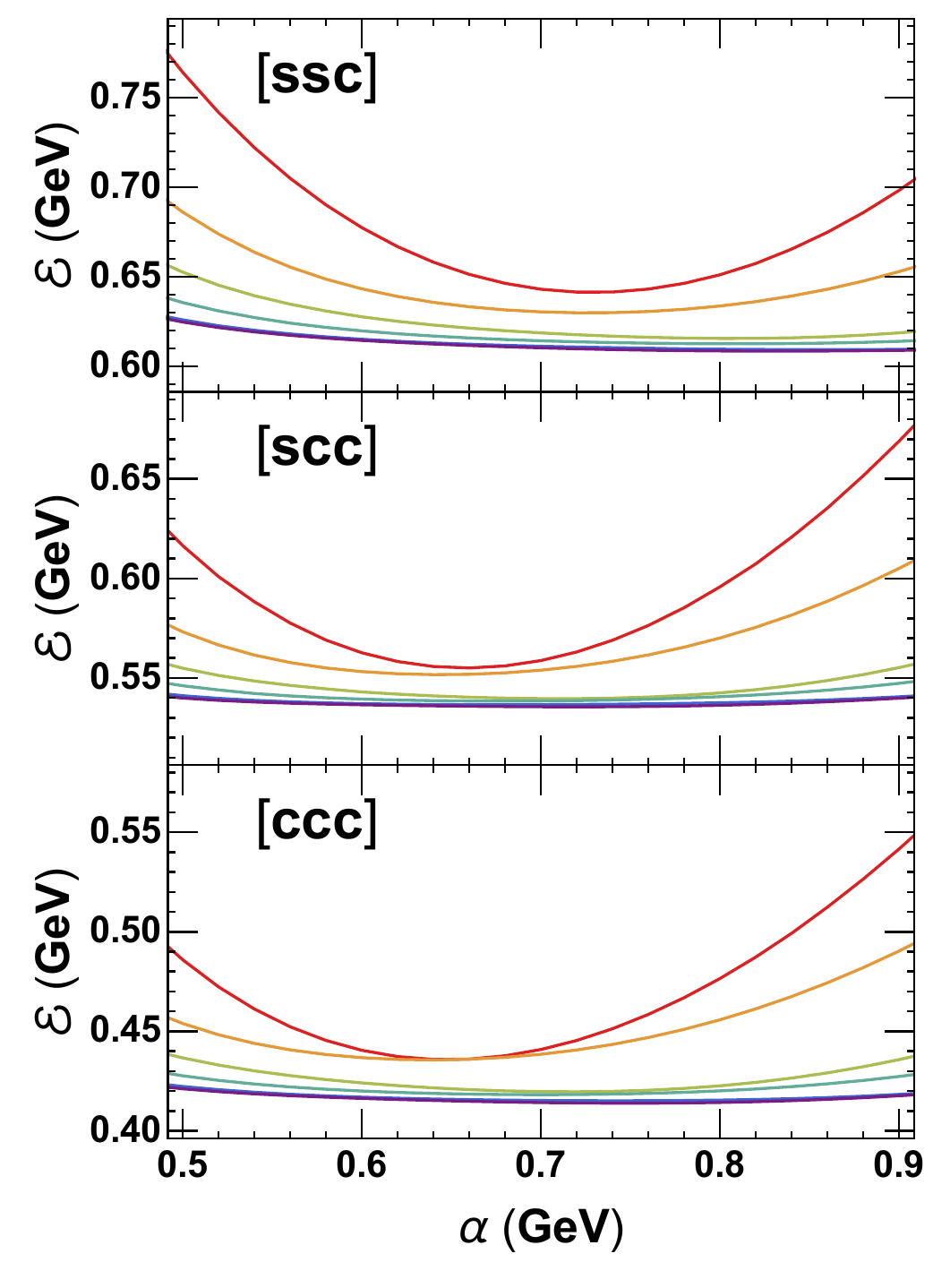}
	\caption{ The binding energy $\mathcal E$ as a function of the scaling parameter $\alpha$ for singly, doubly and triply charmed baryon ground states $\left|ssc\right>$, $\left|scc\right>$ and $\left|ccc\right>$. The six curves from top to bottom correspond to the maximal principal quantum number $N_{\max}= 0, 2, 4, 6, 8$ and $10$.
		The two curves with $N_{\max}=8$ and $10$ almost coincide.
		\label{fig1}}
\end{figure}

The number of the oscillator states or the size of the Hilbert space is controlled by the total principal quantum number $N=2n_\rho+2n_\lambda+l_\rho+l_\lambda$. We truncate the Hilbert space by taking $N \leq N_{max} = 0, 2, 4, 6, 8$ and $10$. Considering the degeneracy of the wave functions, the corresponding number of states is $1, 28, 210, 924, 3003$ and $8008$. From the parity symmetry, which decouples states with odd and even $N$, and momentum conservation, which decouples states with different total magnetic quantum number $m_\rho +m_\lambda$, the effective numbers of states are reduced to $1, 8, 34, 108, 259$ and $560$. We first consider three specific baryon ground states $\left|ssc\right>$, $\left|scc\right>$ and $\left|ccc\right>$ containing one, two and three charm quarks. We calculate their binding energy $\mathcal E=M-\widetilde M$ and check its stability when $N_{max}$ is large enough. The result is shown in Figure \ref{fig1} for the three states and six values of $N_{\max}$. With increasing $N_{\max}$, the $\alpha$-dependence of $\mathcal E$ becomes more and more smooth and reaches almost a constant for $N_{\max}=8$ and $10$. For the state $\left|ssc\right>$, the minimum binding energy obtained from the variational method is $\mathcal E=0.642,\ 0.630,\ 0.616,\ 0.613,\ 0.610$ and $0.609$ GeV corresponding to $\alpha=0.72,\ 0.72,\ 0.80,\ 0.80,\ 0.84$ and $0.84$ GeV, when $N_{\max}$ increases from $0$ to $10$. For the other two states $\left|scc\right>$ and $\left|ccc\right>$, the $\alpha$ dependence is similar, but the binding energy reaches stability much faster. In the following calculation, we will take $N_{\max}=10$ which already guarantees the stability of the numerical calculation.

%%%%%%%%%%%%%%%%%%%%%%%%%%%%%%%%%%%%%%%%%%%%%%%%%%%%%%%%%%%%%%%%%%%%%%%%%%%%%%
%%%%%%%%%%%%%%%%%%%%%%%%%%%%%%%%%%%%%%%%%%%%%%%%%%%%%%%%%%%%%%%%%%%%%%%%%%%%%%

To self-consistently describe both the meson and baryon states, we take a universal set of parameters, by fitting the heavy flavor meson and baryon masses. The parameters we used, including the vacuum quark masses and coupling strengths $\alpha$ and $\sigma$, are shown in Table \ref{table1}. Note that, the relations between the couplings $\alpha_{qq}=0.45\,\alpha_{q\bar q}$ and $\sigma_{qq}=0.49\,\sigma_{q\bar q}$ are close to the results from one gluon exchange and lattice QCD.  

%%%%%%%%%%%%%%%%%%%%%%%%%%%%%%%%%%%%%%%%%%%%%%%%%%%%%%%%%%%%%%%%%%%%%%
\begin{table}[!hbt]
\centering
\begin{tabular*}{2in}{@{\extracolsep{\fill}}lllll}
\hline\hline
&$m_u=m_d=0.135$ GeV &\\
&$m_s=0.263$ GeV &\\
&$m_c=1.400$ GeV &\\
&$m_b=4.773$ GeV &\\
&$\alpha_{qq}=\alpha_{q\bar q}/2.22 = 0.20$ &\\
&$\sigma_{qq}=\sigma_{q\bar q}/2.04 = 0.09$ GeV$^2$ &\\
\hline\hline
\end{tabular*}
\caption{The universal set of parameters of the potential model.
\label{table1}}
\end{table}
%%%%%%%%%%%%%%%%%%%%%%%%%%%%%%%%%%%%%%%%%%%%%%%%%%%%%%%%%%%%%%%%%%%%%%%%

%%%%%%%%%%%%%%%%%%%%%%%%%%%%%%%%%%%%%%%%%%%%%%%%%%%%%%%%%%%%%%%%%%%%%%%%
\begin{table}[!hbt]
\centering
\begin{tabular*}{2.5in}{@{\extracolsep{\fill}}llccc}
\hline
\hline
Meson\qquad & $J^P$& $M_m^E$ & $M_m^T$& $D_R$ \\
	\qquad & & (GeV) & (GeV) &(\%) \\
\hline
$D^{0}$ & $0^-$ & 1.865 & 1.940 & $4.0$ \\
$D^{*0}$& $1^-$ & 2.007 & 2.066 & $3.0$ \\
$D^{+}$ & $0^-$ & 1.870 & 1.940 & $3.8$ \\
$D^{*+}$& $1^-$ & 2.010 & 2.066 & $2.8$ \\
$D_{s}$&  $0^-$ & 1.968 & 2.028 & $3.1$ \\
$D_{s}^*$& $1^-$ & 2.112 & 2.157 & $2.1$ \\
\hline
$\eta_{c}$ & $0^-$& 2.984 & 2.990 & $0.2$ \\
$\eta_{c}(2S)$& $0^-$ & 3.637 & 3.609 & $0.8$ \\
$h_{c1}$& $1^+$& 3.525 & 3.506 & $0.5$ \\
$J/\psi$& $1^-$ & 3.097 & 3.123 & $0.8$ \\
$\psi(2S)$ & $1^-$& 3.686 & 3.701 & $0.4$ \\
$\chi_{c0}$ & $0^+$& 3.415 & 3.442 & $0.8$ \\
$\chi_{c1}$& $1^+$ & 3.511 & 3.504 & $0.2$ \\
$\chi_{c2}$& $2^+$ & 3.556 & 3.519 & $1.0$ \\
\hline
$B^{-}$ & $0^-$ & ~5.279 & ~5.326 & $0.5$ \\
$B^{-*}$& $1^-$ & ~5.325 & ~5.371 & $0.9$ \\
$B^{0}$ & $0^-$ & ~5.280 & ~5.326 & $0.9$ \\
$B^{0*}$& $1^-$ & ~5.325 & ~5.371 & $0.9$ \\
$B_{s}$& $0^-$ & ~5.367 & ~5.408 & $0.8$ \\
$B_{s}^*$& $1^-$ & ~5.415 & ~5.458 & $0.8$ \\
\hline
$\eta_{b}$ & $0^-$ & ~9.399 & ~9.378 & $0.2$ \\
$\eta_{b}(2S)$& $0^-$ & ~9.999 & ~9.964 & $0.3$ \\
$h_{b1}$& $1^+$ & ~9.899 & ~9.918 & $0.2$ \\
$\Upsilon(1S)$ & $1^-$ & ~9.460 & ~9.507 & $0.5$ \\
$\Upsilon(2S)$ & $1^-$ & 10.023 & 10.025 & $0.0$ \\
$\chi_{b0}$ & $0^+$ & ~9.859 & ~9.878 & $0.2$ \\
$\chi_{b1}$& $1^+$ & ~9.893 & ~9.912 & $0.2$ \\
$\chi_{b2}$& $2^+$ & ~9.912 & ~9.929 & $0.2$ \\
\hline
\hline
\end{tabular*}
\caption{The experimentally measured~\cite{Tanabashi:2018oca} and model calculated heavy flavor meson masses $M_E$ and $M_T$. $D_R=|(M_T-M_E)/M_E|$ is the relative difference.
\label{table2}}
\end{table}
%%%%%%%%%%%%%%%%%%%%%%%%%%%%%%%%%%%%%%%%%%%%%%%%%%%%%%%%%%%%%%%%%%%%%%%%%

The calculated heavy flavor meson mass $M_T$ and the comparison with the experimentally measured mass $M_E$~\cite{Tanabashi:2018oca} are shown in Table \ref{table2}. From the relative difference $D_R=|(M_T-M_E)/M_E|$ between the data and model calculation, the two-body Dirac equation describes heavy flavor mesons reasonably well, especially for charmonia and all bottomed mesons with relative difference $D_R < 1\%$. Since the potential description of the interaction between a light and a heavy quark is no longer a good enough approximation, the relative difference $D_R$ becomes larger for $D$ mesons, $D_R\sim (2-4)\%$. 

%%%%%%%%%%%%%%%%%%%%%%%%%%%%%%%%%%%%%%%%%%%%%%%%%%%%%%%%%%%%%%%%%%%%%%%%%%%%
\begin{table}[!hbt]
\centering
\begin{tabular*}{2.5in}{@{\extracolsep{\fill}}llccc}
\hline
\hline
Baryon\qquad & $J^P$& $M_b^E$ & $M_b^T$ & $D_R$ \\
	\qquad & & (GeV) & (GeV) & (\%)    \\
\hline
$\Lambda_c^{+}$ &${(1/2)}^+$ & 2.286	& 2.440	& $6.8$	 \\
$\Sigma_c^{++}$&${(1/2)}^+$ & 2.454	& 2.413	& $1.6$	 \\
$\Sigma_c^{+}$&${(1/2)}^+$ & 2.453	& 2.413	& $1.5$	 \\
$\Sigma_c^{0}$&${(1/2)}^+$ & 2.454	& 2.413	& $1.6$ 	 \\
$\Xi_c^{+}$& ${(1/2)}^+$ & 2.468	& 2.557	& $3.6$	 \\
$\Xi_c^{0}$& ${(1/2)}^+$ & 2.471	& 2.557	& $3.5$	 \\
$\Xi_c^{'+}$& ${(1/2)}^+$ & 2.577	& 2.566	& $0.4$  	 \\
$\Xi_c^{'0}$& ${(1/2)}^+$ & 2.579	& 2.566	& $0.5$  	 \\
$\Omega_c^{0}$&${(1/2)}^+$ & 2.695	& 2.681	& $0.5$	 \\
$\Xi_{cc}^{++}$& ${(1/2)}^+$ & 3.621	& 3.632	& $0.3$	 \\
$\Xi_{cc}^{+}$& ${(1/2)}^+$ & 3.619 	& 3.632	& $0.4$ 	 \\
$\Omega_{cc}^{+}$&${(1/2)}^+$ & 	 	& 3.745	& 			 \\
$\Sigma_c^{*++}$&${(3/2)}^+$ & 2.518	& 2.429	& $3.6$	 \\
$\Sigma_c^{*+}$&${(3/2)}^+$ & 2.518	& 2.429	& $3.6$	 	 \\
$\Sigma_c^{*0}$&${(3/2)}^+$ & 2.518	& 2.429	& $3.6$	 	 \\
$\Xi_c^{*+}$& ${(3/2)}^+$ & 2.646 	& 2.567	& $3.0$ 		 \\
$\Xi_c^{*0}$& ${(3/2)}^+$ & 2.646	& 2.567	& $3.0$	  	 \\
$\Omega_c^{*0}$&${(3/2)}^+$ & 2.766	& 2.689	& $2.8$		 \\
$\Xi_{cc}^{*++}$& ${(3/2)}^+$ & ~		& 3.644	& 	  	 \\
$\Xi_{cc}^{*+}$& ${(3/2)}^+$ & ~		& 3.644	& 	  	 \\
$\Omega_{cc}^{*+}$&${(3/2)}^+$ & ~		& 3.754	& 	  	 \\
$\Omega_{ccc}^{++}$&${(3/2)}^+$ & ~		& 4.784	& 	 	 \\
\hline
$\Lambda_b^{0}$ &${(1/2)}^+$ & 5.620	& 5.793	& $3.1$	 	 \\
$\Sigma_b^{+}$&${(1/2)}^+$ & 5.811	& 5.769	& $0.7$	 	 \\
$\Sigma_b^{0}$&${(1/2)}^+$ 	& 		& 5.769	& 	  	 \\
$\Sigma_b^{-}$&${(1/2)}^+$ 	& 5.816	& 5.769	& $0.8$		 \\
$\Xi_b^{0}$& ${(1/2)}^+$ & 5.792	& 5.913	& $2.1$	 \\
$\Xi_b^{-}$& ${(1/2)}^+$ & 5.795	& 5.913	& $2.0$	  	 \\
$\Xi_b^{'0}$& ${(1/2)}^+$ & 5.792	& 5.903	& $1.9$	 \\
$\Xi_b^{'-}$& ${(1/2)}^+$ & 5.795	& 5.903	& $1.9$	  	 \\
$\Omega_b^{-}$&${(1/2)}^+$ & 6.046	& 6.021	& $0.4$		 \\
$\Xi_{bb}^{-}$& ${(1/2)}^+$ &  		& 10.210	& 	  	 \\
$\Xi_{bb}^{0}$& ${(1/2)}^+$ &  		& 10.210	& 	  	 \\
$\Omega_{bb}^{-}$&${(1/2)}^+$ &  		& 10.319	& 	  	 \\
$\Sigma_b^{*+}$&${(3/2)}^+$ & 5.832 	& 5.781	& $0.9$	 	 \\
$\Sigma_b^{*0}$&${(3/2)}^+$ &  		& 5.7	81	& 	  	 \\
$\Sigma_b^{*-}$&${(3/2)}^+$ & 5.835 	& 5.781	& $0.9$	 	 \\
$\Xi_b^{*0}$& ${(3/2)}^+$ &  		& 5.915	& 	  	 \\
$\Xi_b^{*-}$& ${(3/2)}^+$ &  		& 5.915	& 	  	 \\
$\Omega_b^{*-}$&${(3/2)}^+$ &  		& 6.033	& 	  	 \\
$\Xi_{bb}^{*+}$& ${(3/2)}^+$ & 		& 10.221	& 	 	 \\
$\Xi_{bb}^{*0}$& ${(3/2)}^+$ & 		& 10.221	& 	 	 \\
$\Omega_{bb}^{*-}$&${(3/2)}^+$ &  		& 10.331	& 	 	 \\
$\Omega_{bbb}^{-}$&${(3/2)}^+$ &  		& 14.499	& 	 	 \\
\hline
\hline
\end{tabular*}
\caption{The experimentally measured~\cite{Tanabashi:2018oca} and model calculated heavy flavor baryon masses $M_E$ and $M_T$. $D_R$ is the relative difference between the data and our calculation.
\label{table3}}
\end{table}
%%%%%%%%%%%%%%%%%%%%%%%%%%%%%%%%%%%%%%%%%%%%%%%%%%%%%%%%%%%%%%%%%%%%%%%%%%

The baryon masses and the comparison with the experimental data are shown in Table~\ref{table3}. For doubly charmed baryons $\Xi_{cc}^+$ and $\Xi_{cc}^{++}$ which have been observed experimentally, the agreement between the calculation and the data is good with relative difference $D_R < 0.4\%$. Theoretically, we expect that the relativistic potential model should be more reliable in describing triply charmed and doubly and triply bottomed baryons. For singly charmed or bottomed baryons, however, the model calculation is not so good, since the two light quarks in such a baryon make the potential approximation no longer good enough. Especially for those baryons with two light quarks $u$ and $d$, the relative difference becomes much larger. For $\Lambda_c^+$ it reaches $\sim 7\%$. It is phenomenologically suggested that~\cite{SilvestreBrac:1996bg}, a three body force $C/(m_1m_2m_3)$ controlled by the constituent quark masses may improve the calculation for heavy flavor baryons, especially for the baryons with two light quarks.

%%%%%%%%%%%%%%%%%%%%%%%%%%%%%%%%%%%%%%%%%%%%%%%%%%%%%%%%%%%%%%%%%%%%%%%%%%
\begin{table}[!hbt]
\centering
\begin{tabular*}{3.0in}{@{\extracolsep{\fill}}l|cc|cc|c}
\hline
\hline
& \multicolumn{2}{c|}{Experiment} & \multicolumn{2}{c|}{Model} &  \\
\cline{2-5}
Baryon & $J^P$& $M_E$ & $J^P$ & $M_T$ & $D_R$ \\
& & (MeV) & & (MeV) &(\%) \\
\hline
$\Omega_c^{0}$& ${(1/2)}^+$	& 2.695	& ${(1/2)}^+$(1S)	& 2.681	& $0.5$	 \\
$\Omega_c^*(2770)^{0}$& ${(3/2)}^+$ 	& 2.766	& ${(3/2)}^+$(1S)	& 2.689	&  $2.8$	 \\
$\Omega_c(3000)^{0}$ & 	& 3.000 	& ${(1/2)}^-$(1P) 	& 2.990	&   $0.3$\\
$\Omega_c(3050)^{0}$& & 3.050 	& ${(3/2)}^-$(1P)	& 3.052	&   $0.1$\\
$\Omega_c(3065)^{0}$ & & 3.065 	& ${(1/2)}^-$(1P)	& 3.074	&   $0.3$\\
$\Omega_c(3090)^{0}$ & & 3.090 	& ${(3/2)}^-$(1P)	& 3.085	&   $0.2$\\
$\Omega_c(3120)^{0}$ & & 3.119 	& ${(5/2)}^-$(1P) 	& 3.252	&   $4.3$\\
\hline
\hline
\end{tabular*}
\caption{The experimentally measured and model calculated ground and excited states of $\Omega_c$.
\label{table4}}
\end{table}
%%%%%%%%%%%%%%%%%%%%%%%%%%%%%%%%%%%%%%%%%%%%%%%%%%%%%%%%%%%%%%%%%%%%%%%

With the expansion method in a complete and orthogonal Hilbert space, we can calculate not only the baryon ground states but also the excited states. Table \ref{table4} shows the result for $\Omega_c[ssc]$. $\Omega_c^0$ and $\Omega_c^*(2770)^0$ are the ground states with $J^P=(1/2)^+$ and $(3/2)^+$, and the other five baryons are excited states. The masses of the excited states have been experimentally measured, but the total spins and parities are not yet fixed. From our model calculation, they are all $1P$ states with $J^P=(1/2)^-, (3/2)^-$ and $(5/2)^-$. The mass difference among the excited states is mainly from the spin-orbit and spin-spin interactions. The expansion method in our calculation guarantees the accuracy of the calculation, the relative difference for the masses is $D_R<4.3\%$.

\section{Heavy flavors in the hot medium }
\label{s4}
%%%%%%%%%%%%%%%%%%%%%%%%%%%%%%%%%%%%%%%%%%%%%%%%%%%%%%%%%%%%%%%%%%%%%
\begin{figure}[!hbt]\centering
	\includegraphics[width=170pt]{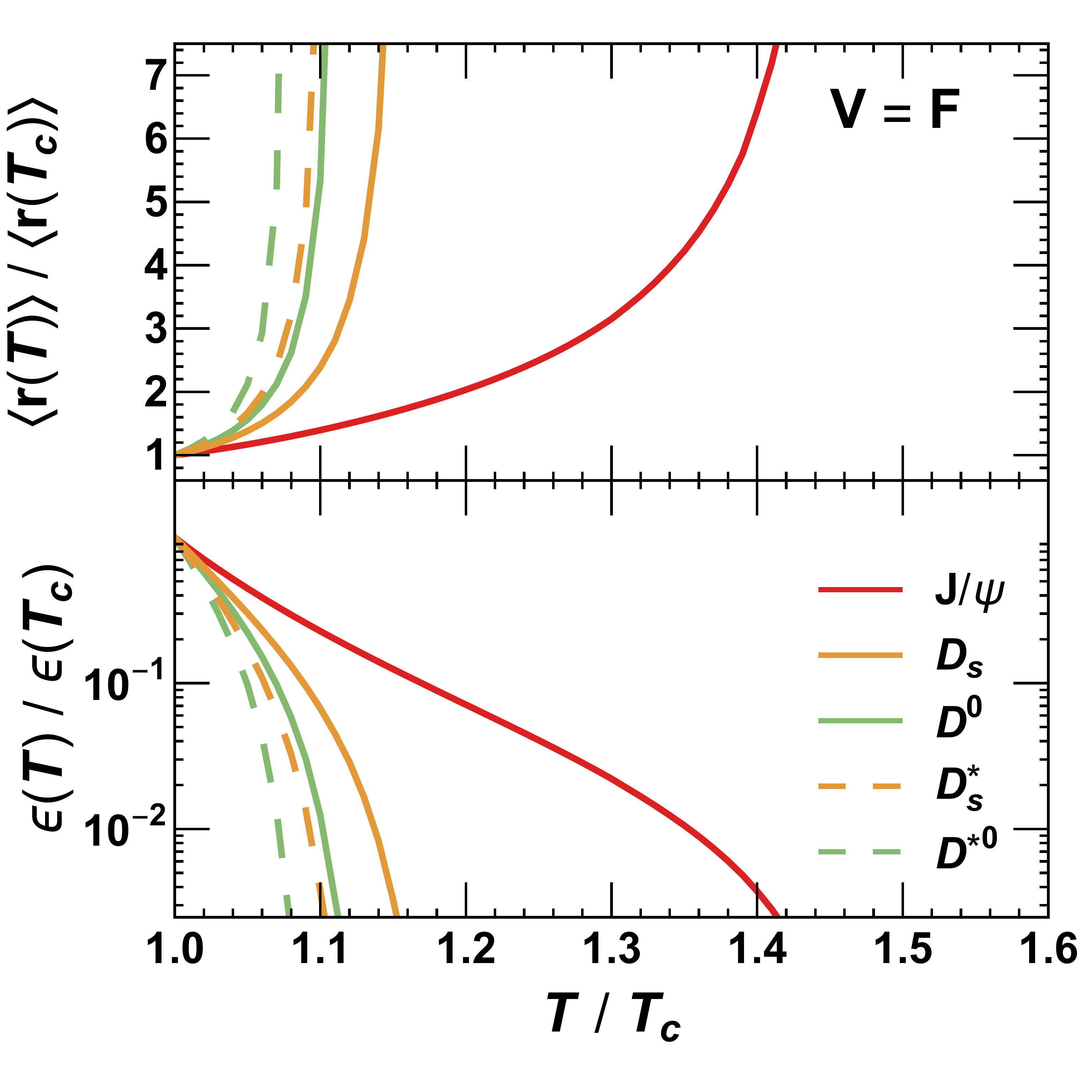}\ ~ \ ~  \includegraphics[width=170pt]{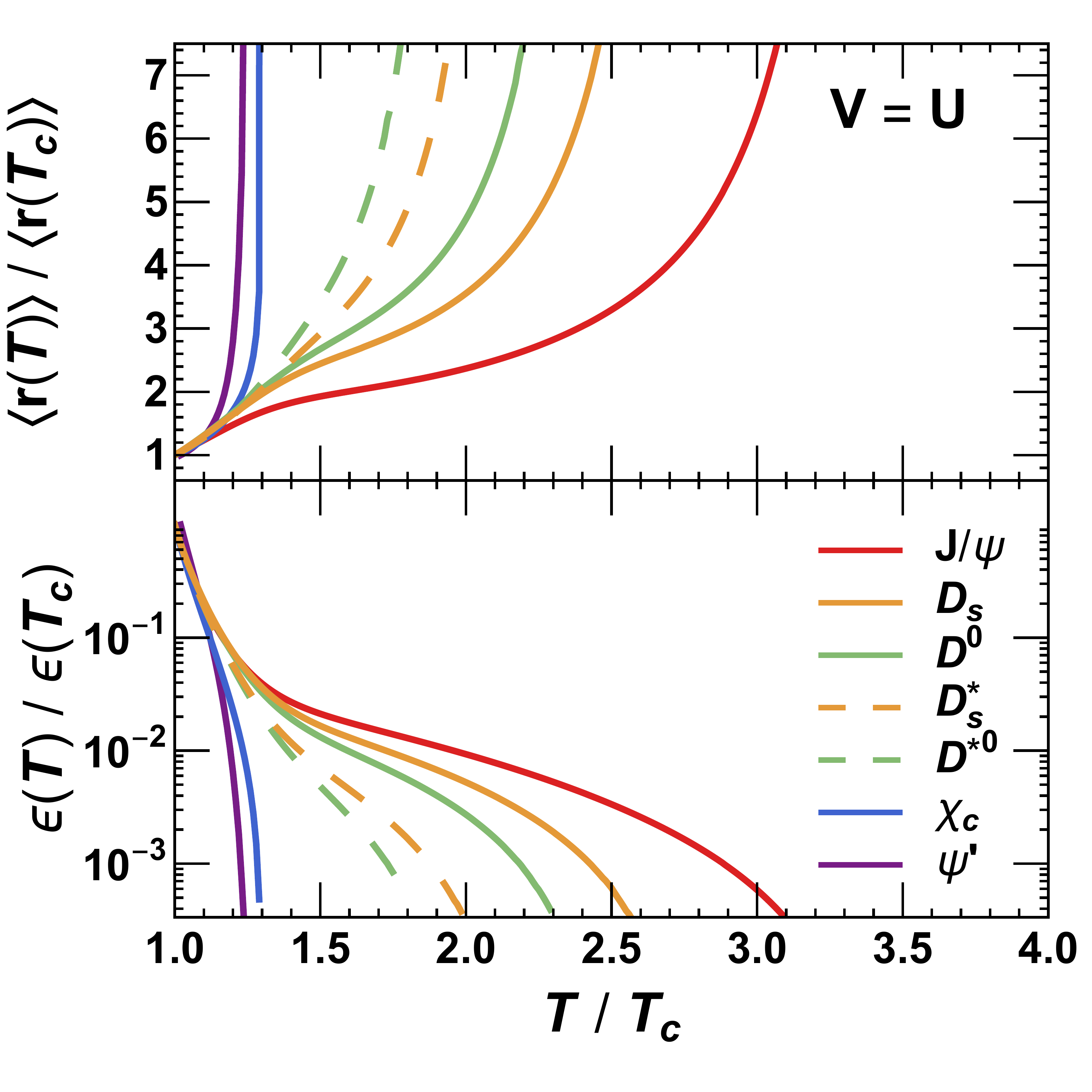}
	\caption{The scaled meson binding energy and root-mean-squared radius as functions of temperature in the two limits of $V=F$ and $V=U$. 
		\label{fig2}}
\end{figure}
%%%%%%%%%%%%%%%%%%%%%%%%%%%%%%%%%%%%%%%%%%%%%%%%%%%%%%%%%%%%%%%%%%%%%%
%%%%%%%%%%%%%%%%%%%%%%%%%%%%%%%%%%%%%%%%%%%%%%%%%%%%%%%%%%%%%%%%%%%%%%
\begin{figure}[!hbt]\centering
	\includegraphics[width=170pt]{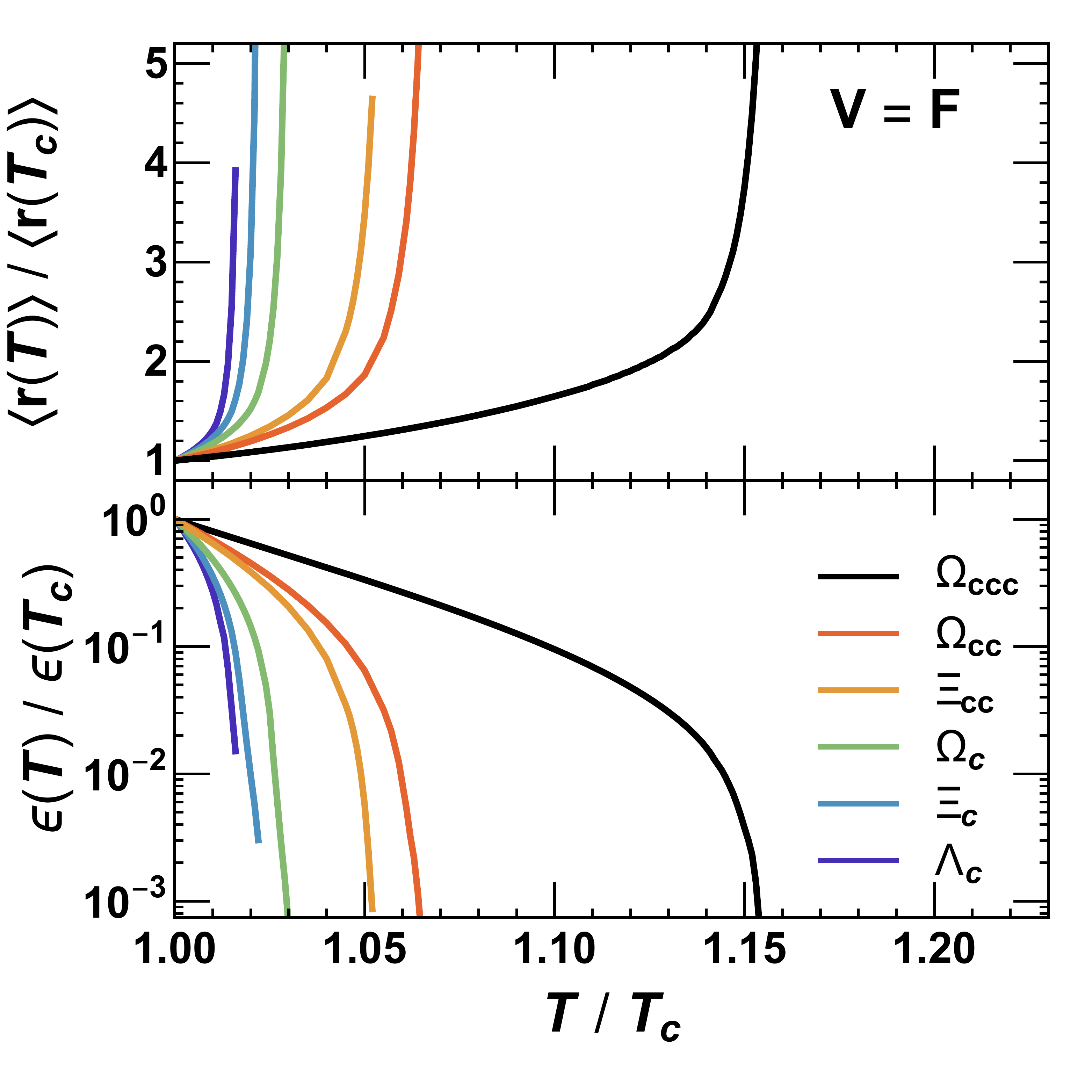} \ ~ \ ~ \includegraphics[width=170pt]{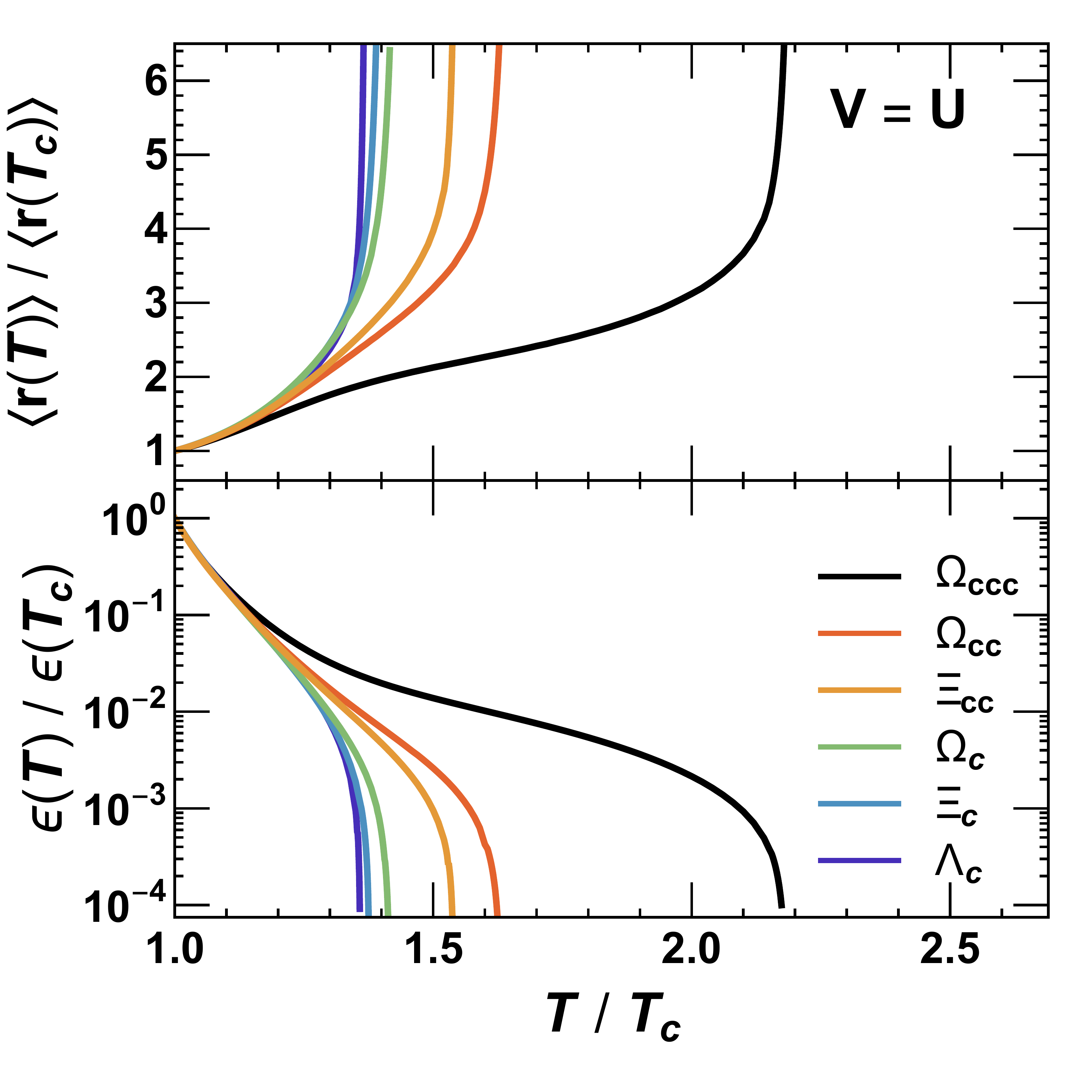}
	\caption{The scaled baryon binding energy and root-mean-squared radius as functions of temperature in the two limits of $V=F$ and $V=U$. 
		\label{fig3}}
\end{figure}
%%%%%%%%%%%%%%%%%%%%%%%%%%%%%%%%%%%%%%%%%%%%%%%%%%%%%%%%%%%%%%%%%%%%%%%
We now turn to study the properties of heavy flavor hadrons in the hot medium. 
Different from light hadrons which are all produced at the end of the quark-gluon plasma, heavy flavor hadrons can be produced in initial hard scattering and survive in the quark-gluon plasma, due to their larger binding energies compared to light hadrons.
However, because of the medium screening effect, the interaction between quarks becomes weaker when temperature increases.
Above certain temperature, the interaction potential well is too narrow to form some (or all) of the bound states, and these hadrons dissociate.
The surviving temperature or the dissociation temperature of a heavy flavor hadron in the hot medium can be determined by solving the corresponding dynamical equation with heavy quark potential at finite temperature. 
This is the original idea of Matsui and Satz~\cite{Matsui:1986dk} who suggested $J/\psi$ suppression as a signal of the quark-gluon plasma formation in heavy ion collisions. 
By solving the non-relativistic Schr\"odinger equation for charmonium states, $\chi_c$, $\psi'$ and $J/\psi$ are found to be sequentially dissociated with increasing temperature~\cite{Satz:2005hx}. 
The potential which controls the absolute value of the dissociation temperature depends on the detail of dissociation process in the hot medium~\cite{Brambilla:2008cx}. 
Yet, its exact form is undetermined on the theoretical side, but can be constrained by two limits. One is the slow dissociation limit, which allows enough time for the constituent quarks to exchange heat with the hot medium and reach thermal equilibrium, and the potential takes the form of the free energy $F$. The other is the rapid dissociation limit, assuming no heat exchange between constituent quarks and the medium, so the potential is the internal energy $U$.
The free energy $F$ has been computed via lattice QCD simulation~\cite{Petreczky:2010yn}, while the internal energy $U$ can be obtained via the thermodynamic relation $U=F+TS$ where $S=-\partial F/\partial T$ is the entropy.
In a general case in high energy nuclear collisions, the potential $V$ is in between these two limits,
$F < V < U$~\cite{Satz:2005hx}.
More discussions on the choice of interaction potential can be found in Refs.~\cite{Liu:2018syc,Lee:2013dca}.
From the above thermodynamic relation one can tell the potential well is deeper in the internal energy limit, hence the dissociation temperature with $V=U$ is higher than that with $V=F$.
Considering the Debye screening at finite temperature, the short and long range interaction potential $V_{q\bar q} = F_{q\bar q} = A_{q\bar q} + S_{q\bar q}$ can be expressed respectively as~\cite{Karsch:1987pv,Satz:2005hx,Shi:2013rga}
\begin{eqnarray}
A_{q\bar q}(r,T)&=&-{\alpha_{q\bar q} \over r}e^{-\mu r}\,, \nonumber\\
S_{q\bar q}(r,T)&=&{\sigma_{q\bar q} \over \mu}\left[{\Gamma(1/4) \over 2^{3/2}\Gamma(3/4)}-{\sqrt{\mu r} \over 2^{3/4}\Gamma(3/4)} K_{1/4}(\mu^2 r^2) \right]\nonumber\\
&&-\alpha_{q\bar q} \mu\,,
\end{eqnarray}
where $\Gamma$ is the Gamma function, $K$ is the modified Bessel function of the second kind, and the temperature dependent screening mass $\mu(T)$ can be extracted by fitting the lattice simulated free energy~\cite{Petreczky:2010yn}. From the known free energy $F$, one can then obtain the other limit of the potential, $V=U$.

From the definition of the dissociation temperature $T_d$ for a heavy quark bound state in hot medium, it is determined by the vanishing binding energy or infinite averaged radius,
\begin{eqnarray}
\epsilon(T_d) &=& 0,\nonumber\\
\langle r(T_d)\rangle &=& \infty.
\end{eqnarray}  

Considering the plateau structure of the heavy quark potential at large distance, see the lattice simulated free energy~\cite{Petreczky:2010yn}, the binding energy $\epsilon$ is the difference between the hadron masses calculated with potential $V(r=\infty,T)$ and $V(r,T)$,
\begin{equation}
\epsilon(T)=M(\infty, T)-M(T).
\end{equation}
For heavy flavor mesons, $M(T)$ is calculated from the two-body Dirac equation and $M(\infty,T)$ is simply expressed as 
\begin{equation}
M(\infty,T) = V_{q\bar q}(\infty, T)+\sqrt{V_{q\bar q}^2(\infty, T)+(m_1+m_2)^2}.
\end{equation}
For heavy flavor baryons, $M(T)$ is calculated from the three-body Dirac equation, and $M(\infty,T)$ is determined by the upper limit of the binding energy, 
\begin{equation}
{1\over 6}\sum_{ij}{\epsilon_j^2 - m_j^2\over \epsilon_i} = \sum_{i<j}\frac{\epsilon_i+\epsilon_j}{2\epsilon_i\epsilon_j} {\cal V}_{ij}(\infty,T)
\end{equation}
with
\begin{equation}
{\cal V}_{ij}(\infty,T) = {2m_im_j\over \epsilon_i+\epsilon_j} V_{qq}(\infty,T) + V_{qq}^2(\infty,T)
\end{equation}
and
\begin{equation}
\epsilon_i = {M(\infty, T) \over 3} + {1\over 3}\sum_{j\neq i}{m_i^2-m_j^2\over \epsilon_i+\epsilon_j}.
\end{equation}

%%%%%%%%%%%%%%%%%%%%%%%%%%%%%%%%%%%%%%%%%%%%%%%%%%%%%%%%%%%%%%%%%%%%%%%
\begin{table}
	\centering
	\begin{tabular*}{2.8in}{@{\extracolsep{\fill}}lccccccc}
		\hline
		\hline
		& $J/\psi $ & $\chi_c$  & $\psi'$  & $D_s$ & $D_s^*$ & $D^0$  & $D^{*0}$\\
		\hline
		$V=F$& 1.42 & - & - & 1.14 & 1.10 & 1.10 & 1.08 \\
		$V=U$& 3.09 &  1.30 & 1.24 & 2.50 & 1.98 & 2.35 & 1.80 \\
		\hline
		\hline
		& $\Omega_{ccc} $ & $\Omega_{cc}$  & $\Xi_{cc}$  & $\Omega_c$ & $\Xi_c$ & $\Lambda_c$\\
		\hline
		$V=F$& 1.15 & 1.06 & 1.05 & 1.03 & 1.02 & 1.02\\
		$V=U$& 2.18 &  1.63 & 1.54 & 1.41 & 1.39 & 1.37\\
		\hline
		\hline
	\end{tabular*}
	\caption{The scaled dissociation temperatures $T_d/T_c$ for charmed mesons and baryons in the two limits of $V=F$ and $V=U$.  
		\label{table5}}
\end{table}
%%%%%%%%%%%%%%%%%%%%%%%%%%%%%%%%%%%%%%%%%%%%%%%%%%%%%%%%%%%%%%%%%%%%%
The scaled meson and baryon binding energy and averaged radius in the two limits of $V=F$ and $V=U$ are shown in Figures~\ref{fig2} and ~\ref{fig3} as functions of scaled temperature. From the recent lattice simulation~\cite{Bazavov:2011nk}, the critical temperature for the QCD  deconfinement phase transition is $T_c\simeq 155$ MeV. With increasing temperature, the binding energy decreases and the averaged radius increases, which indicate a more and more loosely bounded state. Finally, at the dissociation temperature $T_d$ the binding energy reaches zero and the averaged radius becomes infinity, which means the vanishing of the bound state. Since heavy flavors are in more deeply bound states in the limit of $V=U$, they can survive at higher temperatures in this case. The surviving temperatures for charmed mesons and baryons are listed in Table~\ref{table5}. It is clear to see a sequential dissociation for both mesons and baryons, due to their different binding energies.   

\section{Summary}\label{s5}
We investigated heavy flavor hadrons in vacuum and hot medium in the framework of multi-body Dirac equations. With the improved method to solve the three-body Dirac equation, we first fixed a universe set of model parameters by fitting the heavy flavor meson and baryon masses in vacuum. Taking the free energy and internal energy simulated by lattice QCD as the two limits of the heavy quark potential, we then solved the eigenvalues and eigenstates of the two- and three-body Dirac equations at finite temperature, from which we systematically determined the binding energies, averaged sizes and in turn dissociation temperatures for all the heavy flavors. Mesons and baryons are separately sequentially dissociated in the quark-gluon plasma. In comparison with mesons, baryons are easier to be melted in hot medium due to the weaker interaction between quark-quark pairs  than that between quark-antiquark pairs. Considering the limitation of the potential model, the Dirac equations should be systematically improved to better approximate QCD and then used to make better predictions for comparison with the experimental data, especially for mesons with one light quark and baryons with two light quarks. 

{\bf Acknowledgement:} We thank Mr. Tiecheng Guo for the collaboration in the beginning of the work and Profs. Charles Gale and Sangyong Jeon for helpful discussions. JZ and PZ are supported by the NSFC grant Nos. 11575093 and 11890712, and SS is grateful to the Natural Sciences and Engineering Research Council of Canada. Computations were partially made on the supercomputer Bel\'uga, managed by Calcul Qu\'ebec and Compute Canada.

\begin{widetext}
\begin{appendix}
\section{Hamiltonian matrix elements}
\label{aa}
Let's first consider the elements of the baryon kinetic energy ($\sim({\bf \nabla}_\rho^2+{\bf \nabla}_\lambda^2)$) and averaged radius ($\sim (\rho^2+\lambda^2)$), since they can be directly calculated without rotation from $({\boldsymbol \rho},{\boldsymbol \lambda})$ to $({\widetilde {\boldsymbol \rho}},{\widetilde {\boldsymbol \lambda}})$. Taking into account the orthogonality of the spherical harmonic oscillator states, $\left< n_\rho' l_\rho' m_\rho' n_\lambda' l_\lambda' m_\lambda'| n_\rho l_\rho m_\rho n_\lambda l_\lambda m_\lambda \right>= \delta_{n_\rho}^{n_\rho'}\,\delta_{l_\rho}^{l_\rho'}\,\delta_{m_\rho}^{m_\rho'}\, \delta_{n_\lambda}^{n_\lambda'}\,\delta_{l_\lambda}^{l_\lambda'}\,\delta_{m_\lambda}^{m_\lambda'}$, we have 
\begin{eqnarray}
&& -\left< n_\rho' l_\rho' m_\rho' n_\lambda' l_\lambda' m_\lambda'|\nabla_a^2|n_\rho l_\rho m_\rho n_\lambda l_\lambda m_\lambda \right>\\
&=& \delta_{l_\rho}^{l_\rho'}\,\delta_{m_\rho}^{m_\rho'}\,\delta_{n_\lambda}^{n_\lambda'}\,\delta_{l_\lambda}^{l_\lambda'}\,\delta_{m_\lambda}^{m_\lambda'}\alpha^2
	\left[(2n_a+l_a+3/2)\delta_{n_a}^{n_a'}+\sqrt{n_a(n_a+l_a+1/2)}\delta_{n_a}^{n_a'+1}+\sqrt{(n_a+1)(n_a+l_a+3/2)}\delta_{n_a}^{n_a'-1}\right],\nonumber\\
&&\left< n_\rho' l_\rho' m_\rho' n_\lambda' l_\lambda' m_\lambda'|a^2|n_\rho l_\rho m_\rho n_\lambda l_\lambda m_\lambda \right>\nonumber\\
&=& \delta_{l_\rho}^{l_\rho'}\,\delta_{m_\rho}^{m_\rho'}\,\delta_{n_\lambda}^{n_\lambda'}\,\delta_{l_\lambda}^{l_\lambda'}\,\delta_{m_\lambda}^{m_\lambda'}\alpha^{-2}
	\left[(2n_a+l_a+3/2)\delta_{n_a}^{n_a'}-\sqrt{n_a(n_a+l_a+1/2)}\delta_{n_a}^{n_a'+1}-\sqrt{(n_a+1)(n_a+l_a+3/2)}\delta_{n_a}^{n_a'-1}\right].\nonumber
\end{eqnarray}
for $a=\rho, \lambda$. Note that, the kinetic energy and averaged radius are diagonal in spin and flavor spaces.

For the potential ${\cal V}_{ij}({\bf r}_{ij})$ between quark $i$ and quark $j$, the spin-independent terms and spin-spin coupling term depend only on the distance between two quarks. For $\Phi(|{\bf r}_{12}|)$, we have simply
\begin{equation}
\left< n_\rho' l_\rho' m_\rho' n_\lambda' l_\lambda' m_\lambda' |\Phi(|{\bf r}_{12}|)| n_\rho l_\rho m_\rho n_\lambda l_\lambda m_\lambda \right>
= \delta_{l_\rho}^{l_\rho'}\,\delta_{m_\rho}^{m_\rho'}\,\delta_{n_\lambda}^{n_\lambda'}\,\delta_{l_\lambda}^{l_\lambda'}\,\delta_{m_\lambda}^{m_\lambda'}\int \rho^2 \mathrm{d}\rho\, \psi_{n_\rho l_\rho}(\rho) \psi_{n_\rho' l_\rho'}(\rho)\Phi\left(\sqrt{{\overline m(\epsilon_1+\epsilon_2)\over \epsilon_1\epsilon_2}}\rho\right).
\end{equation}
After the rotation (\ref{rotation}), we have similar matrix elements for $\Phi(|{\bf r}_{13}|)$ and $\Phi(|{\bf r}_{23}|)$ with the replacement of $({\boldsymbol \rho},{\boldsymbol \lambda})$ by $({\widetilde {\boldsymbol \rho}},{\widetilde {\boldsymbol \lambda}})$. 

The orbital angular momentum dependent terms and tensor term in ${\cal V}_{ij}({\bf r}_{ij})$ depend on the azimuthal angles $\theta$ and $\phi$ of ${\bf r}_{ij}$. Note that, the couplings $\Phi_{SO}, \Phi_{SOT},  \Phi_{SOD}, \Phi_{SOX}$ and $\Phi_T$ depend only on the distance $|\hat{\bf r}_{ij}|$, the angle dependence comes from their coefficients $H({\boldsymbol \sigma}_i, {\boldsymbol \sigma}_j, {\bf L}_{ij}, \hat{\bf r}_{ij})$ shown in (\ref{potential}). To calculate the matrix elements of the coefficients in flavor and spin spaces, we first construct all possible symmetric, antisymmetric, mixed symmetric and mixed antisymmetric flavor states $\left|F\right>$,
\begin{eqnarray}
\left|F\right>_{S/A} &=& \left[\left|q_1q_2q_3\right>\pm \left|q_2q_1q_3\right> + \left|q_2q_3q_1\right>\pm\left|q_3q_2q_1\right> + \left|q_3q_1q_2\right>\pm\left|q_1q_3q_2\right>\right]/\sqrt 6,\nonumber\\
\left|F\right>_{M1S/A} &=& \left[-2\left|q_1q_2q_3\right> \mp 2\left|q_2q_1q_3\right>+\left|q_2q_3q_1\right>\pm\left|q_3q_2q_1\right> + \left|q_3q_1q_2\right> \pm \left|q_1q_3q_2\right>\right]/\sqrt {12},\nonumber\\
\left|F\right>_{M2S/A} &=& \left[\left|q_2q_3q_1\right>\pm \left|q_3q_2q_1\right> - \left|q_3q_1q_2\right>\mp\left|q_1q_3q_2\right> \right]/\sqrt 4,
\end{eqnarray}
where the symmetry of $\left|F\right>_{M1S}$ and $\left|F\right>_{M2A}$ and antisymmetry of $\left|F\right>_{M2S}$ and $\left|F\right>_{M1A}$ are in the sense of the flavor exchange between the first two quarks $q_1\leftrightarrow q_2$, corresponding to isospin triplet and singlet for $q_1, q_2=u, d$. We then express all possible three-body spin states $\left|S\right>$ as product of two-body spin state $\left|S,S_z\right>_{ij}$ and single spin state $\left|s_z\right>_k$,
\begin{eqnarray}
\left|S\right>_{S1/2} &=& \left|\pm 1/2,\pm 1/2,\pm 1/2\right>=\left|1,\pm 1\right>_{12}\left|\pm 1/2\right>_3 = \left|1,\pm 1\right>_{23}\left|\pm 1/2\right>_1 = \left|1,\pm 1\right>_{31}\left|\pm 1/2\right>_2, \nonumber\\
\left|S\right>_{S3/4} &=& \left[\left|\pm 1/2,\pm 1/2,\mp 1/2\right>+\left|\pm 1/2,\mp 1/2,\pm 1/2\right>+\left|\mp 1/2,\pm 1/2,\pm 1/2\right>\right]/\sqrt 3\nonumber\\
&=& \sqrt {1/3}\left|1,\pm 1\right>_{12}\left|\mp 1/2\right>_3+\sqrt{2/3}\left|1,0\right>_{12}\left|\pm 1/2\right>_3\nonumber\\
&=& \sqrt {1/3}\left|1,\pm 1\right>_{23}\left|\mp 1/2\right>_1+\sqrt{2/3}\left|1,0\right>_{23}\left|\pm 1/2\right>_1\nonumber\\
&=& \sqrt {1/3}\left|1,\pm 1\right>_{13}\left|\mp 1/2\right>_2+\sqrt{2/3}\left|1,0\right>_{13}\left|\pm 1/2\right>_2,\nonumber\\
\left|S\right>_{MS1/2} &=& \left[2\left|\pm 1/2,\pm 1/2,\mp 1/2\right>-\left|\pm 1/2,\mp 1/2,\pm 1/2\right>-\left|\mp 1/2,\pm 1/2,\pm 1/2\right>\right]/\sqrt 6\nonumber\\
&=&\sqrt{2/3}\left|1,\pm 1\right>_{12}\left|\mp 1/2\right>_3-\sqrt{1/3}\left|1,0\right>_{12}\left|\pm 1/2\right>_3\nonumber \\
&=&\sqrt{3/4}\left|0,0\right>_{23}\left|\pm 1/2\right>_1-\sqrt{1/12}\left|1,0\right>_{23}\left|\pm 1/2\right>_1+\sqrt{1/6}\left|1,\pm 1\right>_{23}\left|\mp 1/2\right>_1\nonumber\\
&=&-\sqrt{3/4}\left|0,0\right>_{13}\left|\pm 1/2\right>_2-\sqrt{1/12}\left|1,0\right>_{13}\left|\pm 1/2\right>_2+\sqrt{1/6}\left|1,\pm 1\right>_{13}\left|\mp 1/2\right>_2,\nonumber\\
\left|S\right>_{MA1/2} &=& \left[\left|\pm 1/2,\mp 1/2,\pm 1/2\right>-\left|\mp 1/2,\pm 1/2,\pm 1/2\right>\right]/\sqrt 2\nonumber\\
&=& \left|0,0\right>_{12}\left|\pm 1/2\right>_3\nonumber \\
&=& -{1/2}\left|0,0\right>_{23}\left|\pm 1/2\right>_1 + {1/2}\left|1,0\right>_{23}\left|\pm 1/2\right>_1-\sqrt{1/2}\left|1,\pm 1\right>_{23}\left|\mp 1/2\right>_1\nonumber\\
&=& -{1/2}\left|0,0\right>_{13}\left|\pm 1/2\right>_2 - {1/2}\left|1,0\right>_{13}\left|\pm 1/2\right>_2+\sqrt{1/2}\left|1,\pm 1\right>_{13}\left|\mp 1/2\right>_2,
\end{eqnarray}
where one can clearly see that, the mixed antisymmetric states correspond to the spin singlet of the first two quarks, and the symmetric and mixed symmetric states correspond to all the triplets.

Employing the algebraic method used in quantum mechanics for the single spin operator ${\bf s}_i$ and orbital angular momentum ${\bf L}_{ij}$, we introduce the ladder operators for the relative coordinate $\hat{\bf r}_{ij}$, $\hat r_z$ and ${\hat r}^\pm = ( {\hat r}_x \pm i {\hat r}_y)/\sqrt 2=\sin\theta e^{\pm i\phi}/\sqrt 2$ with the raising and lowering relations,
\begin{eqnarray}
{\hat r}_z \left|l,m\right> &=& \sqrt{{(l+l'+1)^2 - 4m^2\over 4(2l+1)(2l'+1)}}\left(\delta_{l'}^{l+1} + \delta_{l'}^{l-1}\right)\left|l',m\right>,\nonumber\\
{\hat r}^\pm \left|l,m\right> &=& \mp\left[\sqrt{\frac{(l+1\pm m)(l'+1\pm m)}{2(2l+1)(2l'+1)}}\delta_{l'}^{l+1}- \sqrt{\frac{(l \mp m)(l' \mp m)}{2(2l+1)(2l'+1)}}\delta_{l'}^{l-1}\right]\left|l',m\pm1\right>. 
\end{eqnarray}

With the above preparation we are now ready to compute the matrix elements of the coefficients $H(\boldsymbol{\sigma_i},\boldsymbol{\sigma_j},{\bf L}_{ij},{\hat {\bf r}}_{ij})$. For a fixed spin state $\left|S,S_z\right>\left|s_z\right>$ and a fixed spherical harmonic oscillator $\left|n_\rho l_\rho m_\rho n_\lambda l_\lambda m_\lambda\right>$, the matrix elements $H_{LS,L'S'}=\left<l'_\rho m'_\rho\right|\left<S',S'_z\right|H({\boldsymbol \sigma}_i,{\boldsymbol \sigma}_j,{\bf L}_{ij},{\hat {\bf r}}_{ij})\left|S,S_z\right>\left|l_\rho m_\rho\right>$ in angular momentum space can be analytically calculated. They are
\begin{equation}
H_{LS,L'S'}^{SS} = \left[2S(S+1)-3\right]\delta_{l_\rho}^{l'_\rho}\delta_{m_\rho}^{m'_\rho}\delta_S^{S'}\delta_{S_z}^{S'_z}
\end{equation}
for spin-spin coupling with $H={\boldsymbol \sigma}_i\cdot{\boldsymbol \sigma}_j$,
\begin{equation}
H_{LS,L'S'}^{SO} = \left[2 m_\rho S_z \delta_{m_\rho}^{m'_\rho} \delta_{S_z}^{S'_z}+D_{l_\rho}^{m_\rho} D_{S'}^{S'_z} \delta_{m_\rho+1}^{m'_\rho}\delta_{S_z-1}^{S'_z}+D_{l'_\rho}^{m'_\rho} D_S^{S_z} \delta_{m_\rho-1}^{m'_\rho}\delta_{S_z+1}^{S'_z}\right]\delta_{l_\rho}^{l'_\rho} \delta_S^{S'}
\end{equation}
with $D_l^m=\sqrt{(l+m+1)(l-m)}$ for spin-orbital coupling with $H={\bf L}_{ij}\cdot({\boldsymbol \sigma}_i + {\boldsymbol \sigma}_j)$,
\begin{eqnarray}
H_{LS,L'S'}^{SOD/X} &=&  \Big[2 m_\rho \left((1-S)\delta_{S+1}^{S'} \pm (S^2-S_z^2)\delta_{S-1}^{S'}\right)\delta_{m_\rho}^{m'_\rho}\delta_{S_z}^{S'_z}\nonumber\\
&&+\sum_\pm \pm\sqrt{(S\mp S_z-1)(S\mp S_z-2)} D_{l_\rho}^{m_\rho} \left((1-S)\delta_{S+1}^{S'} \mp S\delta_{S-1}^{S'}\right)\delta_{m_\rho\pm 1}^{m'_\rho}\delta_{S_z\mp 1}^{S'_z}\Big]\delta_{l_\rho}^{l'_\rho}
\end{eqnarray}
for spin-orbital difference and cross couplings with $H={\bf L}_{ij}\cdot({\boldsymbol \sigma}_i - {\boldsymbol \sigma}_j)$ and $H=i{\bf L}_{ij}\cdot({\boldsymbol \sigma}_i \times {\boldsymbol \sigma}_j)$,
\begin{eqnarray}
&&H_{LS,L'S'}^T\nonumber\\
&=& \delta_{S}^{S'}\delta_{S_z}^{S'_z}\delta_{m_\rho}^{m'_\rho}\left[6S_z^2-2S(S+1)\right] \left[\frac{l_\rho^2+l_\rho-3m_\rho^2}{4l_\rho^2+4l_\rho-3}\delta_{l_\rho}^{l'_\rho}
	+\frac{3}{2} \sqrt{\frac{((l_\rho+l'_\rho)^2-4m_\rho^2)((l_\rho+l'_\rho+2)^2-4m_\rho^2)}{16(2l_\rho+1)(2l'_\rho+1)(l_\rho+l'_\rho+1)^2}}\left(\delta_{l_\rho+2}^{l'_\rho}
    +\delta_{l_\rho-2}^{l'_\rho}\right)\right]\nonumber\\
&&  -\sum_{\pm}\Bigg\{\sqrt{1\over 2}\delta_S^{S'}\delta^{S'_z}_{S_z\mp 1} \delta_{m_\rho\pm 1}^{m'_\rho}(S\pm S_z)(2S\mp 3S_z)\bigg[ \sqrt{\frac{(l_\rho \mp m_\rho+1)\Gamma(l_\rho \pm m_\rho+4) /\Gamma(l_\rho \pm m_\rho+1)}{(2l_\rho+1)(2l'_\rho+1)(l_\rho+l'_\rho+1)^2}}
	\delta_{l_\rho+2}^{l'_\rho} \nonumber\\
&&	+\sqrt{\frac{(l_\rho \pm m_\rho)\Gamma(l_\rho \mp m_\rho+1) /\Gamma(l_\rho \mp m_\rho-2)}{(2l_\rho+1)(2l'_\rho+1)(l_\rho+l'_\rho+1)^2}}
	 \delta_{l_\rho-2}^{l'_\rho}\pm \frac{(m_\rho+m'_\rho)\sqrt{(l_\rho \pm m_\rho+1)(l_\rho \mp m_\rho)}}{4l_\rho^2+4l_\rho-3}\delta_{l_\rho}^{l'_\rho}\bigg]\nonumber\\
&& +\delta_{S}^{S'}\delta^{S'_z}_{S_z\mp 2} \delta_{m_\rho\pm 2}^{m'_\rho}\sqrt{2(S\pm S_z)(S\pm S_z-1)} \bigg[\frac{\sqrt{((l_\rho+1)^2-(\pm m_\rho+1)^2)(l_\rho^2-(\pm m_\rho+1)^2)}}{4l_\rho^2+4l_\rho-3}\delta_{l_\rho}^{l'_\rho}\nonumber\\
&& +\sqrt{\frac{\Gamma(l_\rho \pm m_\rho+5) /\Gamma(l_\rho \pm m_\rho+1)}{4(2l_\rho+1)(2l'_\rho+1)(l_\rho+l'_\rho+1)^2}} \delta_{l_\rho+2}^{l'_\rho}
   -\sqrt{\frac{\Gamma(l_\rho \mp m_\rho+1) /\Gamma(l_\rho \mp m_\rho-3)}{4(2l_\rho+1)(2l'_\rho+1)(l_\rho+l'_\rho+1)^2}} \delta_{l_\rho-2}^{l'_\rho}
	\bigg]\Bigg\}
\end{eqnarray}
for tensor coupling with $H=3({\boldsymbol \sigma}_i\cdot {\hat{\bf r}}_{ij})({\boldsymbol \sigma}_j\cdot {\hat{\bf r}}_{ij})-{\boldsymbol \sigma}_i\cdot {\boldsymbol \sigma}_j$, and
\begin{equation}
H_{LS,L'S'}^{SOT} ={1\over 3}\sum \left(H_{L'S',L^{''}S^{''}}^T+H_{L'S',L^{''}S^{''}}^{SS}\right)H_{L^{''}S^{''},LS}^{SO}
\end{equation}
for spin-orbital tensor coupling with $H=({\boldsymbol \sigma}_i\cdot{\hat {\bf r}}_{ij})({\boldsymbol \sigma}_j\cdot{\hat {\bf r}}_{ij})({\bf L}_{ij}\cdot({\boldsymbol \sigma}_i+ {\boldsymbol \sigma}_j))$.
\end{appendix}
\end{widetext}

\end{document}